
\input harvmac
\input tables
\def\Title#1#2{\rightline{#1}\ifx\answ\bigans\nopagenumbers\pageno0\vskip1in
\else\pageno1\vskip.8in\fi \centerline{\titlefont #2}\vskip .5in}

%
%
%
\font\ticp=cmcsc10

\def\ajou#1&#2(#3){\ \sl#1\bf#2\rm(19#3)}
\def\jou#1&#2(#3){\unskip, \sl#1\bf#2\rm(19#3)}

\def\frac#1#2{{#1 \over #2}}
\def\eg{{\it e.g.}}
%
%
\lref\ILSe{K. Intriligator, R.G. Leigh, and N. Seiberg, ``Exact
superpotentials  in four dimensions,'' hep-th/9403198\jou
Phys. Rev. &D50 (94) 1092.}
\lref\ADS{I. Affleck, M. Dine, and N. Seiberg, ``Dynamical supersymmetry
 breaking in
supersymmetric QCD,''\ajou Nucl.Phys.& B241 (84) 493\semi
``Dynamical supersymmetry breaking in
four-dimensions and its phenomenological
implications,''\ajou Nucl.Phys. &B256 (85) 557.}
\lref\slan{R. Slansky, ``Group theory for unified model building,''\ajou
Phys. Rep. &79 (81) 1.}
\lref\KeVa{T.W. Kephart and M.T. Vaughn, ``Tensor methods for the
exceptional group $E_6$,''\ajou Ann. Phys. &145 (83) 162.}
\lref\Raca{G. Racah\jou Lincei Rend. Sci. Fis. Mat. Nat. &8 (50) 108.}
\lref\SeibPAS{N. Seiberg, ``The power of duality -- exact results in 4D
SUSY field  theory,''  Rutgers/IAS preprint RU-95-37=IASSNS-HEP-95/46,
hep-th/9506077, to appear in the proceedings of PASCOS 95 and of the Oskar
Klein lectures.}
\lref\SeibER{N. Seiberg, ``Exact results on the space of vacua of four
dimensional SUSY gauge theories,'' hep-th/9402044\jou Phys. Rev. &D49 (94)
6857.}
\lref\Seibdual{N. Seiberg, ``Electric - magnetic duality in supersymmetric
nonabelian gauge theories,''  hep-th/9411149\jou Nucl.Phys. &B435 (95)
129.}
\lref\BDFL{R.E. Behrends, J. Dreitlein, C. Fronsdal, and W. Lee, ``Simple
groups and strong interaction symmetries,''\ajou Rev. Mod. Phys. &34 (62)
1.}
\lref\Cvit{P. Cvitanovi\'c, ``Group theory for Feynman diagrams in
non-abelian gauge theories,''\ajou Phys. Rev. & D14 (76) 1536.}
\lref\Wilsb{V. Novikov, M. Shifman, M. Voloshin, and V. Zakharov,
``Exact Gell-Mann-Low function of
supersymmetric Yang-Mills theories from
instanton calculus,''\ajou Nucl.Phys. &B229 (83) 381\semi
M.A. Shifman and A.I. Vainshtein, ``Solution of the anomaly puzzle in
 SUSY gauge
theories and the Wilson operator expansion,''\ajou Nucl.Phys. &B277 (86)
456.}
\lref\CoDi{S. Cordes and M. Dine, ``Chiral symmetry breaking in
supersymmetric  O(N) gauge theories,''\ajou Nucl.Phys. &B273 (86) 581.}
\lref\InSe{K. Intriligator and N. Seiberg, ``Duality, monopoles, dyons,
confinement  and
oblique confinement in supersymmetric SO($N_c$) gauge theories,'' Rutgers
preprint RU-95-03, hep-th/9503179, to appear in {\sl Nucl. Phys. B}.}
\lref\InPo{K. Intriligator and P Pouliot, ``Exact superpotentials, quantum
vacua and duality in supersymmetric SP($N_c$) gauge theories,'' Rutgers
preprint RU-95-23, hep-th/9505006.}
\lref\Kut{D. Kutasov, ``A comment on duality in N=1
supersymmetric non-abelian gauge theories,'' EFI--95--11,
hep-th/9503086.}
\lref\KuSc{D. Kutasov and A. Schwimmer, ``On duality in
supersymmetric Yang-Mills theory,'' EFI--95--20, WIS/4/95,
hep-th/9505004.}
\lref\ASY{O. Aharony, J. Sonnenschein, and S. Yankielowicz,
``Flows and duality symmetries in N=1 supersymmetric gauge
theories,'' TAUP--2246--95, CERN-TH/95--91, hep-th/9504113.}
\lref\Berk{M. Berkooz, ``The dual of supersymmetric SU(2k)
with an antisymmetric tensor and composite dualities,'' RU-95-20,
hep-th/9505088.}
\lref\ILS{K. Intriligator, R.G. Leigh, and M.J. Strassler, ``New examples
of duality in chiral and non-chiral supersymmetric gauge theories,''
RU-95-38, hep-th/9506148}
\lref\Pesa{I. Pesando, ``Exact results for the supersymmetric $G_2$ gauge
theories,'' NORDITA-95/42P, hep-th/9506139.}
\lref\Int{K. Intriligator, ``New RG Fixed Points and
Duality in Supersymmetric $Sp(N_c)$ and $SO(N_c)$ Gauge
Theories,'' RU--95--27, hep-th/9505051, to appear in {\sl Nucl. Phys.
B}.}
\lref\LeSt{R.G. Leigh and M.J. Strassler, ``Duality of
$Sp(2N_c)$ and $SO(N_c)$ Supersymmetric Gauge Theories with
Adjoint Matter,'' RU--95--30, hep-th/9505088.}
\lref\DRSW{M. Dine, R. Rohm, N.
Seiberg, E. Witten, ``Gluino condensation in superstring models,''\ajou
Phys.Lett. &156B (85) 55.}
\lref\DIN{J.P. Derendinger, L.E. Ibanez, H.P. Nilles, ``On the low-energy
D = 4, N=1 supergravity theory
extracted from the D = 10, N=1 superstring,''\ajou Phys.Lett. &155B (85) 65.}
\lref\DKLP{L. Dixon, V. Kaplunovsky, J. Louis, and M. Peskin,
unpublished\semi reported in L. Dixon, ``Supersymmetry breaking in string
theory,'' In {\sl The Rice meeting: proceedings},
proceedings of the 1990 APS DPF Conference, Houston TX Jan
3-6, 1990, B. Bonner and H. Miettinen, eds. (World Scientific 1990).}
\lref\Marc{J. March-Russell, private communication.}
\lref\LuTa{M.A. Luty and W. Taylor IV, ``Varieties of vacua in classical
supersymmetric gauge theories,'' MIT preprint MIT-CTP-2440,
hep-th/9506098.}
\lref\Intrint{K. Intriligator, `` `Integrating in' and exact superpotentials
in 4-d,'' hep-th/9407106\jou Phys.Lett. &B336 (94) 409.}
\noblackbox
\Title{\vbox{\baselineskip12pt\hbox{UCSBTH-95-14}
\hbox{hep-th/9506196}
}}
{\vbox{\centerline {Some Exact Results in Supersymmetric }
\vskip2pt\centerline{Theories Based on Exceptional Groups\ }
}}

\centerline{{\ticp Steven B. Giddings}\footnote{$^\dagger$}
{Email address: giddings@denali.physics.ucsb.edu} and
{\ticp John M. Pierre}\footnote{$^*$}
{Email address: jpierre@sbphy.physics.ucsb.edu}  }
\vskip.1in
\centerline{\sl Department of Physics}
\centerline{\sl University of California}
\centerline{\sl Santa Barbara, CA 93106-9530}
\bigskip
\centerline{\bf Abstract}

We begin an investigation of supersymmetric theories based on exceptional
groups.  The flat directions are most easily parameterized using their
correspondence with gauge invariant polynomials.  Symmetries and holomorphy
tightly constrain the superpotentials, but due to multiple gauge invariants
other techniques are needed for their full determination.  We give an
explicit treatment of $G_2$ and find gaugino condensation for $N_f\leq 2$,
and an instanton generated superpotential for $N_f=3$.  The analogy with
$SU(N_c)$ gauge theories continues with modified and unmodified quantum
moduli spaces for $N_f=4$ and $N_f=5$ respectively, and a non-Abelian
Coulomb phase for $N_f\geq6$.  Electric variables suffice to describe this
phase over the full range of $N_f$.  The appendix gives a self-contained
introduction to $G_2$ and its invariant tensors.

\Date{}

\newsec{Introduction}

Recent exact results have
reinvigorated the study
of supersymmetric gauge theories.\foot{For a recent
review with complete references see \refs{\SeibPAS}.}
These results follow from applying
the powerful constraints of symmetry and
holomorphy of the superpotential.  They are interesting for several reasons.
First, our understanding of how nature could be described by a theory with
spontaneously broken supersymmetry is far from complete.  A better
understanding of such
models with dynamical supersymmetry breaking should yield
further insight into this problem.  Second, until now the strong-coupling
behavior of non-abelian gauge theories has been poorly understood.  The
exact results offer new approaches to confinement and other
interesting features of the vacua of these theories.  Furthermore, the
growing evidence for strong-weak coupling duality suggests the possibility
of opening up a completely new window on strongly coupled theories, through
which much more may be learned about their physical properties.

By applying these tools to a variety of theories one hopes to get a better
understanding of their underpinnings and of the different phenomena that
can occur in supersymmetric gauge theories.  The list of models
studied has been growing, and includes the simple
groups $SU(N)$ \refs{\SeibER,\Seibdual},
$SO(N)$ \refs{\Seibdual,\InSe}, and $SP(2N)$ \refs{\InPo}, with matter in
the fundamental, as well as theories with matter in higher representations
\refs{\Kut\KuSc\Int\LeSt\ASY\Berk-\ILS} and product groups\ILS.
In this paper we initiate a study of the remaining simple groups,
the exceptional groups $G_2$, $F_4$, $E_6$, $E_7$, and $E_8$.

Beyond our interest in expanding the knowledge of supersymmetric theories
and the tools used to study them, there are at least two reasons to
consider the exceptional groups.  First, the largest exceptional group,
$E_8$,
arises in the gauge symmetry $E_8\times E_8$ of the heterotic string.
Furthermore, it has been proposed that non-perturbative effects such as
gaugino condensation in one of the $E_8$ factors\refs{\DIN,\DRSW} or
the racetrack variants\refs{\DKLP} could be the origin of
supersymmetry breaking in string theory.  This motivates us
to better understand such strong-coupling phenomena in $E_8$ and its
subgroups, which include the other exceptional groups.
Second, even if string
theory does not describe the physical world, $E_6$ models have been
seriously considered as possible grand unified theories.  If this group
appears in nature, it could also conceivably
play a role in the supersymmetry breaking
sector.

Study of exceptional groups is more difficult than that of the other simple
groups due to their greater algebraic complexity.  In particular, the
problem of explicitly parameterizing flat directions appears formidable.
An alternate
approach is to use the result that the flat directions correspond
to gauge invariant polynomials.\foot{This result has long been implicit in
the literature; recent proofs of it are \refs{\Marc,\LuTa}.}  One can then
use known results about invariant tensors in the exceptional groups to
attempt to construct all invariant polynomials, and take these as a
starting point for the analysis.
Even this problem is difficult; so far we have only managed to
explicitly treat $G_2$.  Nonetheless, this or other closely related
approaches based on investigating the decomposition of these models under
maximal subgroups should in principle yield an exact treatment of the
remaining exceptional groups.

We begin the next section by investigating some generic features of
exceptional groups.  We give a brief discussion of the problem of
parameterizing vacua, then use symmetries to find general constraints on the
form of the superpotential.  Unfortunately it appears that symmetry
arguments alone are not sufficient to determine the superpotential.  We
then investigate the possible emergence of a non-abelian Coulomb phase for
certain values of $N_f$, the number of flavors.

In section two we give a detailed analysis of the group $G_2$.  We are able
to deduce the form of the superpotential for $N_f\leq 3$ by using knowledge
of the invariant polynomials, symmetry constraints, and the technique of
``integrating in\refs{\Intrint}.''  We find gaugino condensation at
$N_f\leq 2$ and an instanton generated superpotential at $N_f=3$. As in the
case of $SU(N)$, we also find a modified quantum moduli space for
$N_f=4$ and a quantum moduli space equivalent to the classical one at
$N_f=5$.  We then argue that for $N_f\geq6$ the theory should have a
non-abelian Coulomb phase as its infrared description.  A minor novelty is
that the ``electric'' description should be valid all the way down to
$N_f=6$: there is no domain where a magnetic description is mandatory to
describe the dynamics.  This is fortunate, since we have not yet been able
to deduce the dual magnetic theory.

The appendix contains a more or less self-contained treatment of $G_2$.  We
explicitly construct this group in a way that its maximal $SU(3)$ subgroup
is manifest by treating it as the subgroup of $SO(7)$ that leaves a real
spinor invariant.  From this construction we derive the invariant
tensors and the relations among them.

Upon completion of this work, we received \refs{\Pesa}, which arrives at
many of the same results in the $G_2$ theory.

\newsec{General results in exceptional models}

In this section we will make some general observations on models based
on exceptional groups, with matter ``quark'' fields ${Q^i_\alpha}$ in
the defining, or fundamental, representation. Here greek indices are group
indices and latin indices label flavors.  Dimensions of these
representations\refs{\slan}
are shown in table I. With the exception of ${E_6}$
the representations are real. We denote quarks in the anti-fundamental
of ${E_6}$ by ${\bar Q}^{\alpha}_{\tilde\imath}$. In the case of
${E_8}$, the adjoint is the smallest representation and will be used for
the quarks.
\bigskip
\begintable
|$D_F$|$D_A$| $C_F$ | $C_A$ | \rm Primitive invariants \cr
{}~~ | ~~ | ~~ | ~~| ~~ | ~~ \cr
$G_2$| 7 | 14 | 1 | 4 |
${\delta^{\alpha\beta}},{f^{\alpha\beta\gamma}}$\cr
$F_4$|26|52|3|9|${\delta^{\alpha\beta}},{d^{\alpha\beta\gamma}}$\cr

$E_6$|27|78| 3| 12|
${d^{\alpha\beta\gamma}}$\cr
$E_7$ | 56 | 133 | 6| 18 |
${f^{\alpha\beta}},{d^{\alpha\beta\gamma\delta}}$\cr
$E_8$|--  | 248 | -- | 30 |
${\delta^{AB}},{C^{ABC}_3}$,${\cdots}$\endtable
\item{} {\bf Table I.}
Shown are some properties of the exceptional groups.  $D_F$ and
$D_A$ denote the dimensions of the fundamental and adjoint respectively,
and $C_F$ and $C_A$ the Dynkin index of these representations, with
normalization convention
$C_F(SU(N))={1\over 2}$.
\bigskip
Study of these theories requires a parameterization of their D-flat
directions, that is, solutions of the constraint

\eqn\one{{D^A} = {\sum_I}{Q_I^{*\alpha}}{T^{A\beta}_{\alpha}}{Q_\beta^I}=0 }

\noindent where ${T^{A\beta}_{\alpha}}$ are the group generators.
Although explicit parameterizations of these can be given for the
non-exceptional groups \refs{\ADS,\InSe,\InPo}, the
complicated algebraic structure of the exceptional groups makes them
more challenging.  Alternatives are to decompose the exceptional groups
into non-exceptional subgroups, or to use the fact
\refs{\Marc,\LuTa} that flat directions can be parametrized by the gauge
invariant polynomials in the quark fields.  In this paper we adopt the
latter approach.

To form these polynomials we need the invariant tensors in the
fundamental representation. The ``primitive'' tensors from which these
can be constructed are given for groups other than ${E_8}$ in \refs{\Cvit}.
In addition to the tensors ${\delta^\alpha_\beta}$ and
${\epsilon^{\alpha_1\cdots{\alpha_D}_F}}$, which are invariant in all
cases, the exceptional groups have either fully symmetric primitive
tensors, denoted by $d$'s in table I, or totally antisymmetric primitives,
denoted by $f$'s in table I.

For $E_8$, with quarks in the adjoint, the invariant tensors are not
explicitly known.  Two of them are $\delta^{AB}$ and the structure
constants $c^{ABC}$.  There are also independent
Casimirs at orders 8, 12, 14, 18, 20, 24, and 30, which can be used to form
invariants\refs{\Raca}.

To find all gauge invariants, we must construct all independent contracted
products of
these tensors. There are a finite number of independent combinations due
to the existence of relations among products of the primitive tensors.
Some of these are given in \refs{\Cvit}, although the complete set of
these identities is apparently in general not known.

For example, in $E_6$ the primitives are $\delta^\alpha_\beta$,
$d^{\alpha\beta\gamma}$,  $d_{\alpha\beta\gamma}$,
$\epsilon^{\alpha_1\cdots\alpha_{27}}$, and
$\epsilon_{\alpha_1\cdots\alpha_{27}}$.  From these
we can also form invariants\refs\KeVa\
such as $d_{\alpha\beta\gamma}
d^{\gamma\delta\epsilon}$ and $\epsilon^{\alpha_1\cdots\alpha_{27}}
d_{\alpha_{27} \beta\gamma}$.  However, nontrivial relations
such as $d_{\alpha\beta\gamma}d^{\beta\gamma\epsilon} =
10 \delta_\alpha^\epsilon$ and the Springer relation
\eqn\Sreln{d_{\epsilon\phi\eta}\left(d^{\phi\alpha\beta}
d^{\eta\gamma\delta} + d^{\phi\alpha\gamma} d^{\eta\beta\delta}
+d^{\phi\alpha\delta} d^{\eta\beta\gamma}\right) = \delta_\epsilon^\alpha
d^{\beta\gamma\delta} + \delta_\epsilon^\beta d^{\alpha\gamma\delta} +
\delta_\epsilon^\gamma d^{\alpha\beta\delta} +
\delta_\epsilon^\delta d^{\alpha\beta\gamma} }
can be used to reduce many of the higher products.  In the appendix we will
discuss the analogous
problem for $G_2$ in detail.

Note that, with the exception of ${G_2}$, even the one flavor case
always has more than one non-trivial invariant, \eg\ $\,M =
{\delta^{\alpha}_\beta}{Q_\alpha}{{\bar Q}^\beta}$,
${D}={d^{\alpha\beta\gamma}}{Q_\alpha}{Q_\beta}{Q_\gamma}$, {\it etc.}
in ${E_6}$.
These invariants parameterize the different subgroups to
which the quark vevs may break the original group. For example, the ${\bf
27}$ can break ${E_6}$
to the distinct maximal subgroups ${SO}({10})$ and ${F_4}$.

Although we will not give a full treatment of the supersymmetric theory
for arbitrary exceptional groups here, some general features of these
theories can be deduced from symmetries. In the next section we will
explicitly treat the group ${G_2}$, and in that case fill in more of the
details.

With ${N_f}$ flavors, the non-chiral theories ${G_2}, {F_4}, {E_7},
{E_8}$ have the classical symmetries ${U}(1)_A$,

\eqn\Axial{ {Q^i_\alpha}{\rightarrow}{e^{i\phi}}{Q^i_\alpha}, }

\noindent ${U}(1)_X$,

\eqn\two{ {Q^i_\alpha}(\theta){\rightarrow}{Q^i_\alpha}
({\theta}{e^{-i\phi}}), }

\noindent and ${SU}({N_f})$. In the ${E_6}$ theory, these extend to
${U}(1)_A$,

\eqn\three { {Q^i_\alpha}{\rightarrow}{e^{i\phi}}{Q^i_\alpha},\hskip
.25truein {\bar
Q}^{\alpha}_{\tilde\imath}{\rightarrow}{e^{i\phi}}{\bar
Q}^{\alpha}_{\tilde\imath}, }

\noindent ${U}(1)_B,$

\eqn\four { {Q^i_\alpha}{\rightarrow} {e^{i\phi}}{Q^i_\alpha},\hskip
.25truein {\bar
Q}^{\alpha}_{\tilde\imath}{\rightarrow} {e^{-i\phi}}{\bar
Q}^{\alpha}_{\tilde\imath}, }

\noindent and ${SU}({N_f}){\times}{SU}({N_f})$, as well as ${U}(1)_X$.
The non-anomalous R symmetries are given by
\eqn\five{ {J^\mu_X} - \frac{C_A-n_FC_F}{n_F C_F} {J^\mu_A}, }
where ${n_F} = {N_f}$ for ${G_2}, {F_4}, {E_7}$, and
${E_8}$, and ${n_F} = {2N_f}$ for ${E_6}$.

These can be used to constrain the form of the superpotential as with
other groups. To see this, first recall that the exact result for the
Wilsonian ${\beta}$-function gives \refs{\Wilsb}

\eqn\six{ e^{-{8\pi^2\over{g^2(\mu)} } +i\theta} =
\Biggl(\frac{\Lambda}{\mu}\Biggr)
^{3C_A-n_F C_F}, }

\noindent where ${\Lambda}$ is the ${UV}$ cutoff.
Therefore we can treat ${U}(1)_A$ as unbroken if we take
${\Lambda}$ to transform as

\eqn\seven{ {\Lambda^{3C_A-n_F C_F}}{\rightarrow}{e^{2in_F C_F\phi}}{\Lambda^
{3C_A-n_F C_F}} }

\noindent under \Axial. Using this to constrain the superpotential gives
\eqn\eight{ {W} = {f} \Biggl( \frac{Q^{2n_F C_F}}{\Lambda^{3C_A-n_F
C_F}}\Biggr). }

\noindent Demanding that $W$ have charge two under ${U}(1)_R$ then implies

\eqn\nine{ {W} {\sim} \frac{ {\Lambda^{{3C_A-n_F C_F}\over{C_A-n_F C_F}}}}
{Q^{2n_FC_F/(C_A-n_FC_F)}} .}

\noindent However, this together with the ${S}{U}({N_f})$ symmetry is in
general not sufficient to uniquely fix the potential since there can be
more than one invariant with the correct symmetries. ${G_2}$ will
furnish an explicit example of this in the next section.

Note that for the superpotential to be instanton generated, it must be
proportional to $e^{-8\pi^2/g^2}={\Lambda^{3C_A-n_FC_F}}$, i.e.
\eqn\ten{C_A-n_FC_F = 1. }
This is only possible \refs{\CoDi} for ${G_2}$ with ${N_f} =
{3}$.  In the other cases where the superpotential is not instanton
generated, we should nonetheless be able to deduce its form by relating
theories using 1) integrating out/in heavy quarks, which changes the number
of flavors, and 2) allowing quarks to get large vevs, which changes the
size of the group through the Higgs effect.  One might for example be able
to use these techniques to relate the $E_8$ theory without matter
to other theories with
matter fields in the adjoint \refs{\Kut-\ILS} and thus prove gaugino
condensation in $E_8$ theories.

Parallelling the analysis of the non-exceptional groups, notable
theories are those where the superpotential must vanish due to vanishing
of the quark ${R}$ charge,

\eqn\eleven{n_F = \frac{C_A}{C_F}, }

\noindent and those where asymptotic freedom is lost,

\eqn\twelve{ n_F = {3} \frac{C_A}{C_F}. }

As in other theories, it is natural to conjecture the existence of an
interacting non-abelian Coulomb phase for some range of
${n_F}{\leq}\frac{3C_A}{C_F}$. The exact 1PI ${\beta}$-function is
\refs{\Wilsb}

\eqn\obfn{ {\beta_g} =- \frac{g^3}{16\pi^2}
\frac{3C_A-(1-\gamma)C_Fn_F}
{1-C_A {{g^2}\over{8\pi^2}}} }

\noindent where

\eqn\thirteen{ \gamma = -{C_F}\hskip .10truein \frac{D_A}{D_F}\hskip
.10truein \frac{g^2}{4\pi^2}\hskip .10truein + {\cdots} }

\noindent is the anomalous mass dimension, and near ${n_F} =
\frac{3C_A}{C_F}$ cancellation of one-and two-loop terms appears
possible. If there is such a fixed point, described by a superconformal
theory, as argued in \refs{\Seibdual} the chiral operators should have
dimensions

\eqn\exdim{ D = \frac{3}{2}{|R|}, }

\noindent where ${R}$ is the ${R}$ charge of the operator. Gauge invariant
meson
operators can be formed in all of the exceptional models, and by
\exdim\ they have dimension

\eqn\fourteen{ D(QQ) = {3} \frac{n_FC_F-C_A}{n_FC_F}. }

\noindent This value agrees with that given by ${\gamma}+{2}$, if one
assumes vanishing of the ${\beta}$-function \obfn.

The unitarity constraint ${D}({Q}{Q}) \geq {1}$ tells us where
this hypothesized ``electric'' description must fail, below

\eqn\belec{ n_F= \frac{3}{2}\hskip .10truein \frac{C_A}{C_F}. }

\noindent At this value the meson is a free field.

In non-exceptional theories, in a range below this ${N_f}$ magnetic
variables are conjectured \refs{\Seibdual} to be necessary to describe
the IR dynamics.
However, the analogous range may not necessarily exist for exceptional
groups. In the next section, we'll see that for ${G_2}$ a description in
terms of composite mesons and baryons appears appropriate at ${N_f} = 5$;
${N_f}= 6$ is the direct analogue of ${N_f} = {N_c} + 2$ in
${SU}({N_c})$, and should give the lower bound on the magnetic
description. However

\eqn\fifteen{ \frac{3}{2}\hskip .10truein \frac{C_A}{C_F} = {6} }

\noindent so ${N_f} = {6}$ also corresponds to the lower bound on the
electric description. There is no range where magnetic variables furnish
the only possible IR description.

In the other exceptional groups, it is natural to conjecture that the
analogue of ${N_f} = {N_c}$ in ${SU}({N_c})$ is

\eqn\sixteen{ n_F = \frac{C_A}{C_F}; }

\noindent here the superpotential must vanish and
one anticipates a quantum moduli space different from the
classical one. (We will show this for ${G_2}$.)  Taking the potentially
dangerous step of pushing the analogy further, in
each case the lower bound or the magnetic description would then be at

\eqn\seventeen{\eqalign{ n_F &= \frac {C_A}{C_F} + 2
\quad(G_2,F_4,E_7,E_8)\ ,\cr
n_F &= \frac {C_A}{C_F} + 4\quad (E_6)\ , }}

\noindent and this exceeds \belec\ for all exceptional groups. Thus an
electric description may be sufficient in each case.

Nonetheless, a magnetic description could add useful insight into the
dynamics of these theories. Even for ${G_2}$ we have not yet found this
description. Some clues exist; for example in all simple non-exceptional
groups the theory is self dual at the value of ${N_f}$ for which an
added adjoint matter field yields a vanishing ${\beta}$-function. For
exceptional groups, this would happen at

\eqn\eighteen{ n_F = \frac{2C_A}{C_F}. }

\noindent We will make further comments on the ${G_2}$ case in the
following section.

\newsec{Supersymmetric ${G_2}$  gauge theory}

\indent The example we consider is an N=1 supersymmetric ${G_2}$  gauge theory
with ${N_f}$ flavors of quarks in the fundamental {\bf 7} representation.
Using the techniques developed in \SeibER,\ILSe\ we obtain exact
results in the quantum theory. We recover features which are similar to
those obtained for ${SU}({N_c}), {SO}({N_c})$ and ${Sp}({N_c})$
gauge theories
with matter in the fundamental representation. This is further
evidence of a set of properties generic to N=1 supersymmetric gauge theories.

\indent In particular we find gluino condensation and instanton generated
superpotentials for ${N_f}{\leq}{2}$ and ${N_f} = {3}$ respectively,
with no ground state in the massless limit. For ${N_f} = {4}$ there is
a moduli space of inequivalent vacua which is smoothed out in the quantum
theory by a one instanton effect. For ${N_f} = {5}$ the classical and
quantum moduli space are the same and there is confinement without
chiral symmetry breaking at the origin. For ${N_f}{\geq}{6}$, we
expect a nonabelian Coulomb phase to describe the infrared physics.

\subsec{Gauge invariant fields}

\indent As discussed in the preceding section, the light degrees of freedom
on the moduli space can be labeled by the ${G_2}$  gauge invariant polynomials
of the fundamental quarks subject to possible constraints.
As shown in the appendix, these polynomials can be constructed from
the three independent invariant tensors ${\delta^{\alpha\beta}}$,
${f^{\alpha\beta\gamma}}$ and ${\tilde f}^{\alpha\beta\gamma\delta}$.
Combining these with the quarks ${Q^i_\alpha}$
leads to the composite fields

\eqn\nineteen{
\eqalign{{M^{ij}}&= {\delta^{\alpha\beta}}{Q^i_\alpha}{Q^i_\beta}\cr
{B}_{{i_4}{\cdots}{i}_{N_f}}& =  {{1}\over{3!}}
{\epsilon}_{{i_1}{i_2}{i_3}{i_4}
{\cdots}{i}_{N_f}}{f^{\alpha\beta\gamma}}{Q^{i_1}_\alpha}{Q^{i_2}_\beta}
{Q^{i_3}_\gamma}\cr
{F}_{{i_5}{\cdots}{i}_{N_f}}& = {{1}\over{4!}}
{\epsilon}_{{i_1}{i_2}{i_3}{i_4}{i_5}{\cdots}{i}_{N_f}}{\tilde
f}^{\alpha\beta\gamma\delta}{Q^{i_1}_\alpha} {Q^{i_2}_\beta}
{Q^{i_3}_\gamma}{Q^{i_4}_\delta}\ .\cr }}

\subsec{Gaugino condensation for ${N_f}{\leq}{2}$}

\indent  Due to the antisymmetry of ${f^{\alpha\beta\gamma}}$ and
${\tilde f}^{\alpha\beta\gamma\delta}$, ${B}$ and
${F}$ vanish for ${N_f}{\leq}{2}$ and the only light fields
are the ${M^{ij}}$.  In addition to holomorphy,  $U(1)_R$ symmetry and
dimensional analysis,
which constrain the form of the superpotential to \nine,
invariance under the ${SU}({N_f})$ global flavor symmetry further
restricts it to be

\eqn\twenty{
{{W_{\rm eff}}} = {{({4}-{N_f}) ({\Lambda}_{N_f})^{{12-N_f}
\over{4-N_f}}}\over
{{\rm det} {M}^{{1}\over{4-N_f}}}}\ . }

\noindent This is exact;  the normalization of ${\Lambda}_{N_f}$ has
been adjusted to set threshold corrections to unity \ILSe.
${G_2}$ has a maximal ${SU}({3})$ subgroup under which ${\bf
7}{\rightarrow}{\bf 3} {+} {\bf {\bar 3}} {+} {\bf 1}$ (see appendix).
Therefore as we turn on vevs, the generic breaking sequence is
\eqn\bseq{ {G_2} {\buildrel N_f=1\over \longrightarrow} {SU}({3})
{\buildrel N_f=2\over\longrightarrow} {SU}({2})
{\buildrel N_f=3\over\longrightarrow} {\phi}. }

\noindent For ${N_f} = {1}$ or ${2}$, gluino condensation will occur
\refs{\SeibER} in the sector with the unbroken
pure ${SU}({4} - {N_f})$ gauge theory
through nonzero expectation values of the gluino bilinears,

\eqn\twentyone{
\langle{\lambda\lambda}\rangle_{4-N_f}
{\sim} {\Lambda^3}_{SU(4-N_f)}\ . }

\noindent Matching the couplings at the ${G_2}$ breaking scale
${v} = ({\rm det}{M})^{1/2N_f}$  relates the scales of the two
theories,

\eqn\twentytwo{
{\Lambda^3}_{SU(4-N_f)} = {v^3} \biggl({{{\Lambda}_{N_f}}\over{v}}\biggr)
^{{12-N_f}\over{4-N_f}}\ . }

\noindent Since $\langle{\lambda\lambda}\rangle_{4-N_f}$ is the coefficient
of the dynamically generated F-term \ADS\ we
see that gluino condensation leads to a term of the same form as \twenty.
Therefore supersymmetry is spontaneously broken but there is no ground state,
as in ${SU}({N_c})$ with ${N_f} < {N_c}  - {1}$.

Adding a tree level mass term ${W_{\rm tree}} = {m_{ij}} {M^{ij}}$
to \twenty\  and solving for ${M^{ij}}$ gives the expectation values
\eqn\mexp{\eqalign{
\langle{M^{ij}}\rangle &= \biggl({\Lambda}_{N_f}\biggr)^{{12-N_f}\over{4}}
({\rm det}{m})^{{1}\over{4}} ({{1}\over{m}})^{ij}\cr
{\rm det} \langle{M^{ij}}\rangle &=
\biggl({\Lambda}_{N_f}\biggr)^{{(12-N_f)N_f}\over{4}}
({\rm det}{m})^{{N_f-4}\over{4}}\ . }}
The symmetries ensure that \mexp\ generalizes to higher values
of ${N_f}$  when ${B} = {F} = {0}$. For ${m^{ij}} {\not=} {0}$
there are four different supersymmetric ground states. Notice\refs{\CoDi}
that this disagrees with the Born-Oppenheimer calculation of the Witten
index\lref{\Witt}, which would give three.

\subsec{Instanton generated superpotential for ${N_f}  = {3}$}

\indent For ${N_f}  = {3}$ the light fields are labeled by the ``mesons''
${M^{ij}} = {\delta^{\alpha\beta}}{Q^i_\alpha}{Q^j_\beta}$
and a ``baryon''  ${B} = {{1}\over{3!}} {\epsilon_{ijk}}
{f^{\alpha\beta\gamma}}
{Q^i_\alpha}{Q^j_\beta}{Q^k_\gamma}$. Holomorphy, symmetries and
dimensions restrict the superpotential to the form

\eqn\twentythree{
{W_{\rm eff}} = {{\Lambda^9_3}\over{{\rm det}{M}}} {f}\biggl( {{B^2}
\over{{\rm det}{M}}}\biggr). }

\noindent Since ${W_{\rm eff}}$ has the right quantum numbers we expect that it
is instanton generated along the flat directions where the gauge symmetry
is completely broken.  However it seems that the form of $f(x)$ cannot be
deduced purely from symmetry arguments.  The exact form can instead be
found using the ``integrating in'' technique of \refs{\Intrint}
by taking ${G_2}$
with ${N_f}  = {3}$
as the ``upstairs'' theory and ${N_f}  = {2}$ as the ``downstairs'' theory.
The dynamically generated superpotential \twenty\ of the downstairs theory is

\eqn\twentyfour{
{W_d} = {{2\Lambda^5_2}\over {\sqrt{{\rm det}{M_d}}}}\ . }

\noindent Since the upstairs theory contains a non-quadratic gauge
invariant there may be additional terms in the downstairs theory of the form
${W_{{\rm tree},{d}}} + {W_{\Delta}}$. By turning on a tree level
superpotential for the ``new" flavor ${Q}$  in the upstairs theory
\eqn\twentyfive{
{W_{{\rm tree},{u}}} = {m}{Q^3}{\cdot}{Q^3} + {bB} }
and integrating out the heavy quark we find
\eqn\twentyfivep{
{W_{{\rm tree},{d}}} = {-} {{b^2}\over{4m}} {\rm det} {M_d}\ , }
where ${M_d}$  are the mesons constructed from the
remaining quarks.
${W_{\Delta}}$ is determined by symmetries to be of the form

\eqn\twentysix{
{W_{\Delta}} = {{b^2}\over{m}} {\rm det}{M_d}\,\,{g} \biggl(
{{{b^2}({\rm det}{M_d})^{{3}\over{2}}}\over {{m} {\Lambda^5_2}}}\biggr)
}

\noindent and the limits ${W_{\Delta}}{\rightarrow} {0}$  for ${m}{\rightarrow}
{\infty}$  and ${\Lambda_2}{\rightarrow}{0}$ restrict ${g}({x})$ to be exactly
zero.  ${\Lambda_2}$  is related to ${\Lambda_3}$  by

\eqn\twentyseven{
( {\Lambda_2})^5 = \sqrt{E} ({\Lambda_3})^{{9}\over{2}} }

\noindent where ${E}$ is the scale where the couplings match.
Symmetries restrict ${E}$ to have the form

\eqn\twentyeight{
{E} = {mh} \biggl (
{{{b^2}{\rm det}{M_d}^{{3}\over{2}}}\over
{{m}{\Lambda^5_2}}}\biggr ) }

\noindent and the limits ${E}{\rightarrow}{m}$ as ${\rm det}{M}{\rightarrow}
{\infty}$ and ${m}{\rightarrow}{\infty}$ imply  that ${h}({x}) = {1}$.
Therefore the exact matching condition is

\eqn\twentynine{
({\Lambda_2})^5 = \sqrt{m} ({\Lambda_3})^{{9}\over{2}}. }

\noindent By combining results \twentyfour, \twentyfivep,
and \twentynine, ${W_u}$  can
be obtained from

\eqn\thirty{
{W_n} = {{2{\sqrt m} {\Lambda}^{{9}\over{2}}_{3}}
\over {\sqrt{det M_d}}}
{-} {{b^2}\over{4m}} {\rm det}
{M_d} {-} {mQ_3\cdot Q_3} {-} {bB} }

\noindent by treating ${m}$ and ${b}$ as fields and integrating them out.
The result is that
\eqn\thirtyone{
{W_u} = {{\Lambda^9_3}\over
{{\rm det}{M}-{B^2}}} }
is the dynamically generated superpotential for ${N_f}  = {3}$.  This
theory has no ground state and is similar to ${SU}({N_c})$ with ${N_f}  =
{N_c} - {1}$.

The singularity at det${M} = {B^2}$  is due to extra massless gluinos
at points where some of the gauge symmetry is unbroken. Although the
generic breaking sequence was given in \bseq, at $N_f=3$
there are also
non-trivial flat directions which leave an ${SU}({2})$
subgroup unbroken.  In the basis of the appendix these are easily seen to
be

\eqn\thirtytwo{
\eqalign{
{\langle}Q^1_\alpha{\rangle}&= {v_1}{\delta_{\alpha 7}}\cr
{\langle}{Q^2_\alpha}{\rangle}&= {v_2} ({\delta_{\alpha 1}} + {\delta_{\alpha
4}})
\cr
{\langle}{Q^3_\alpha}{\rangle}&= {v_3} ({\delta_{\alpha 1}} - {\delta_{\alpha
4}}).
\cr} }

\noindent Along these flat directions one can easily check
${\rm det} {M} = {B^2}$  is satisfied.

\subsec{Modified quantum moduli space for ${N_f}  = {4}$}

\indent For ${N_f}  = {4}$ the light degrees of freedom are labeled by

\eqn\thirtythree{
\eqalign{
{M^{ij}}&= {\delta^{\alpha\beta}}{Q^i_\alpha}{Q^j_\beta}\cr
{B_i}&= {{1}\over{3!}} {\epsilon_{ijk\ell}}{f^{\alpha\beta\gamma}}
{Q^j_\alpha}{Q^k_\beta}{Q^\ell_\gamma}\cr
{F}&= {{1}\over{4!}} {\epsilon_{ijk\ell}}{\tilde
f}^{\alpha\beta\gamma\delta}{Q^i_\alpha}{Q^j_\beta}{Q^k_\gamma}
{Q^\ell_\delta}\ .
\cr } }

\noindent The classical constraint is

\eqn\thirtyfour{
{\rm det} {M} - {F^2} - {B_i}{M^{ij}}{B_j} = {0} }

\noindent which can be seen as a consequence of (A.20) and the Bose
symmetry of the quark fields. The expectation value ${\rm det} {M} =
{\Lambda^8_4}$ from \mexp\ implies
that the classical constraint is modified quantum mechanically to

\eqn\thirtyfive{
{\rm det} {M} - {F^2} - {B_i}{M^{ij}}{B_j} = {\Lambda^8_4} }

\noindent and the singularities are smoothed out by a one instanton effect.
The symmetries do not allow a dynamically generated superpotential,
hence there is a moduli space of inequivalent vacua defined by
\thirtyfive\
and this
is different from the classical moduli space.
This is similar to ${SU}({N_c})$
with ${N_f}  = {N_c}$  flavors.

\indent These results can be independently derived by taking the ${N_f} = {3}$
case as the downstairs theory and integrating in a new flavor.  From
\thirtyone\ we
have the dynamically generated superpotential of the downstairs theory
\eqn\thirtysix{
{W_d} = {{\Lambda^9_3}\over
{{\rm det} {M_d} - {B^2}}}. }
Turning on the tree level superpotential
\eqn\thirtyseven{
{W_{{\rm tree},{u}}} = {m}{Q^4}{\cdot}{Q^4} + {2m_I}{Q^I}{\cdot}{Q^4}
+ {b^I}{B_I} + {fF} }

\noindent and integrating out the massive quark gives

\eqn\thirtyeight{
\eqalign{
{W_{{\rm tree},{d}}} = {-} {{1}\over{4m}} \biggl\lbrack {\rm det} {M_d}
{b^I}({M^{-1}_d})_{IJ}{b^J}&+ {f^2} ({\rm det} {M_d} - {B^2}) + {4b^I}{m_I}
{B}\cr
&+ {4m_I} ({M_d})^{IJ}{m_J}\biggr\rbrack\cr} }

\noindent for the downstairs theory (I,J = 1,...,3).  Taking
${W_{\Delta}}  = {0}$ and the
matching condition  $({\Lambda_3})^9 = {m} ({\Lambda_4})^8$ and
performing the inverse Legendre transformation on the full
superpotential of the downstairs theory leads to

\eqn\thirtynine{
{W_n} = {{m\Lambda^8_4}\over{{\rm det}{M_d} - {B^2}}} +
{W_{{\rm tree},{d}}} - {W_{{\rm tree},{u}}}\ . }

\noindent Integrating out ${m}, {m_{I}}, {b^I},$ and  ${f}$  from ${W_n}$
gives  ${W_u}  = {0}$ for the dynamically generated superpotential as
expected.  In addition the equations of motion ${{{\partial} {W_n}}\over
{{\partial} {m}}} = {{{\partial} {W_n}}\over{{\partial} {m_{I}}}} =
{{{\partial} {W_n}}\over
{{\partial} {b^I}}} = {{{\partial} {W_n}}\over{{\partial} {f}}} = {0}$
lead to the quantum constraint \thirtyfive.

\indent The ${N_f}  = {4}$ theory can be described by the effective
superpotential

\eqn\forty{
{W_{\rm eff}} = {X} ({\rm det} {M} {-} {F^2} {-} {B_i}{M^{ij}}{B_j} {-}
{\Lambda^8_4}) }

\noindent where ${X}$ is a Lagrange multiplier field.  Perturbing
\forty\ by adding a tree level mass term  ${W_t}  = {m_{ij}}  {M^{ij}}$,
the results for ${N_f}<{4}$,  \thirtyone, \twenty\
can recovered by integrating out
one or more massive quarks.

\subsec{Quantum moduli space for ${N_f}  = {5}$}

\indent The light fields and their transformations under the global
${SU}({5}){\times}{U}({1})_R$  flavor symmetry are

\eqn\fortyone{
\eqalign{
{M^{ij}}&= {\delta^{\alpha\beta}}{Q^i_\alpha}{Q^j_\beta}\quad\quad
{}~~~~~~~~~~~~~:
{15_{{2}\over{5}}}\cr
{B_{ij}}&= {{1}\over{3!}} {\epsilon_{ijk\ell m}}{f^{\alpha\beta\gamma}}
{Q^k_\alpha}{Q^\ell_\beta}
{Q^m_\gamma}~~~~~~:{\overline{10}}_{{3}\over{5}}\cr
{F_i}&= {{1}\over{4!}}{\epsilon_{ijk\ell m}}{\tilde
f}^{\alpha\beta\gamma\delta}{Q^j_\alpha}{Q^k_\beta}
{Q^\ell_\gamma}{Q^m_\delta}
{}~:{\bar 5}_{{4}\over{5}}\ .\cr }}

\noindent The classical constraint which follows from (A.20) is

\eqn\fortytwo{
\eqalign{
{F_i}{F_j} {-} {B_{ik}}{M^{k\ell}}{B_{\ell j}} {-} {\rm det} {M}
({M^{-1}})_{ij} = {0}. }}

\noindent The expectation values \mexp\
imply the quantum-mechanical modification of this to
\eqn\fortythree{
{F_i}{F_j} - {B_{ik}}{M^{k\ell}}{B_{\ell j}} {-} {\det} {M}
({M^{-1}})_{ij} = ({\Lambda_5})^7 {m_{ij}}. }
The classical constraints are satisfied in the
${m}{\rightarrow} {0}$ limit,
therefore the moduli space of the
massless quantum theory is that same as the
classical theory.

\indent At the origin it appears that the ${SU}({5}){\times}{U}({1})_R$  chiral
symmetry remains unbroken and that all the components of ${M^{ij}}, {B_{ij}}$
and $F_i$  are massless. The 't Hooft anomaly matching conditions
between the fundamental
fermion fields (quarks which transform like  $7 \times
{\bf 5}_{-{4\over5}}$
and 14 gluinos) and those of the massless spectrum \fortyone\
are satisfied

\eqn\fortyfour{
\eqalign{
{SU}({5})^3\quad\quad&{7d}^{(3)}({5})= {d^{(3)}} ({15}) + {d^{(3)}}
({\overline{10}}) +
{d^{(3)}} ({\bar 5})\cr
{SU}({5})^2 {U}({1})_R\quad\quad&
{7}(-{{4}\over{5}}) {d}^{(2)} (5)=  (-{{3}\over{5}})
{d}^{(2)}({15}) + (-{{2}\over{5}}) {d}^{(2)}  ({\overline{10}}) +
(-{{1}\over{5}})
{d}^{(2)}  ({\bar 5})\cr
{U}({1})^3_R\quad\quad&
 35(-{{4}\over{5}})^3  +  {14}=  {15} (-{{3}\over{5}})^3
+ {{10}} (-{{2}\over{5}})^3  + 5 (-{{1}\over{5}})^3\cr
{U}({1})_R\quad\quad&35({-}{{4}\over{5}}) +
{14}= 15(-{{3}\over{5}}) + {{10}}
(-{{2}\over{5}}) + 5(-{{1}\over{5}}).\cr }}

\noindent There is confinement without chiral symmetry breaking at the origin.
This is similar to ${SU}({N_c})$ with ${N_f}  = {N_c}  + {1}$ flavors.

\indent A low energy effective superpotential which obeys all the symmetries
in the problem is given by

\eqn\ficous{
{W_{\rm eff}} = {{1}\over{\Lambda^7_5}} \lbrack {F_i}{M^{ij}}{F_j}
{-} {{1}\over{2}} {B_{ij}}{M^{jk}} {B_{k\ell}}{M^{\ell i}} {-} {\rm det}
{M} {-} {{1}\over{4}} {\epsilon^{ijk\ell
m}}{B_{ij}}{B_{k\ell}}{F_m}\rbrack }

\noindent where the constant coefficients have been chosen so that the
constraint \fortytwo\ arises from the equation of motion ${{{\partial}
{W_{\rm eff}}}\over{{\partial}{M^{ij}}}} = {0}$.  In addition,
two other constraints are obtained from
\eqn\fortyfive{
{{{\partial}{W_{\rm eff}}}\over{{\partial}{F_m}}} =  {2M^{mi}}{F_i} {-}
{{1}\over{4}} {\epsilon^{ijk\ell m}}{B_{ij}}{B_{k\ell}} = {0}  }
and
\eqn\fortysix{
{{{\partial}{W_{eff}}}\over{{\partial}{B_{mn}}}} = {M^{mi}} {B_{ij}}
{M^{jn}} {-} {{1}\over{2}} {\epsilon^{mnijk}} {B_{ij}}{F_k} = {0}. }
These can also be derived from the identities in the appendix
(and fix the coefficient in the last term of \ficous).
By adding a tree level superpotential  ${W_t}  = {m_{ij}}
{M^{ij}}$ and integrating out the massive fields we find the quantum
constraint \fortythree\ as the equation of motion. For the special case
${W_t}  = {m_{55}}{M_{55}}$, by integrating out the one massive field
we recover the results \thirtyfive, \forty\ of the ${N_f}  = {4}$ theory from
the equations of motion.  Similarly the results for ${N_f} < {4}$ can
be recovered by giving masses to more of the fields.

\subsec{Non-abelian Coulomb phase for ${N_f}{\geq}{6}$}

\indent The analysis of ${G_2}$ has closely paralleled that of ${SU}
({N_c})$, so it's not unreasonable to expect that for ${N_f}{\geq}{6}$,
one finds a non-abelian Coulomb phase analogous to that for ${N_f}{\geq}
{N_c} + {2}$ in ${SU}({N_c})$. Indeed, the arguments of
\refs{\Seibdual} show that such a phase should exist for
${N_f}{\geq}{8}$. To see this, note that allowing two of the flavors to
get vevs breaks ${G_2}$ to ${SU}({2})$, with ${N_f}-{2}$ remaining
flavors. This theory is not asymptotically free for $({N_f} - {2}) {\geq}
{6}$, implying the existence of a non-Abelian Coulomb phase.

As suggested in section one, it is not unreasonable to
expect this non-Abelian electric phase to extend down to ${N_f} = {6}$,
where the meson field would have dimension one and become free, as in
${SU}({N_c})$.

Notice that the coincidence between the lower bound on the
electric description and the lower bound on the non-Abelian Coulomb phase
means that a dual magnetic theory is not necessary to describe the
dynamics. However, the ubiquity of such theories suggests one should be
sought here as well.

A standard procedure is to identify the duals of the baryons
with
the baryons of the dual theory. An added complication here is the
existence of the two types of baryons, ${B}$ and ${F}$. The R-charge
assignments for dual quarks appear to be simplest if ${F}$ is taken to
correspond to the baryon in the dual theory, but one is then faced with
identifying ${B}$ in the dual. The difficulty of interpreting this as a
fundamental field suggests either a more complicated group structure or
the necessity for fields transforming in different representations.
Indeed, the fact that ${G_2}$ can be gotten from ${SO}({7})$ through
breaking by a spinor {\bf 8} vev suggests that a promising route is to
investigate the dual of ${SO}({N})$ theories
with both fundamentals and spinors.  A reasonable conjecture is that the
dual theories are given by $SO(N)$'s with both fundamental and spinor
fields.

\newsec{Conclusion}

Using the result that flat directions are parametrized by gauge-invariant
polynomials, one may extend the exact treatment of supersymmetric gauge
theories to the exceptional groups.  Symmetries and holomorphy provide
stringent constraints on the superpotential, but are not sufficient to
fully determine it as there are multiple invariants that can be formed
with the correct transformation properties.  Nonetheless, other techniques
such as ``integrating in'' can be used to obtain the superpotential.  A
necessary first step is to determine the algebraically independent
gauge-invariant polynomials, and this requires knowledge of the group's
invariant tensors and of the relations among them.

This approach has been explicitly used for the group $G_2$.  Gluino
condensation was found for $N_f\leq2$, and an instanton generated
superpotential for $N_f=3$.  At higher $N_f$ the theory also parallels the
$SU(N_c)$ case:  there is a modified quantum moduli space at $N_f=4$, a
moduli space equivalent to the classical one at $N_f=5$, and apparently a
non-abelian Coulomb phase for $N_f\geq6$.  The dual magnetic description
has not been found, but an electric description suffices to treat the full
range of $N_f$.

\bigskip\bigskip\centerline{{\bf Acknowledgements}}\nobreak

We wish to thank S. Chaudhuri, J. March-Russell,
and N. Seiberg for useful discussions, and M.
Bowick and P. Cvitanovi\'c for pointing us to \refs{\Cvit}.
This work was supported in part by
DOE grant DOE-91ER40618 and
by NSF PYI grant PHY-9157463.

\appendix{A}{}

\indent In this Appendix we will give a self contained derivation of
features of ${G_2}$ that are needed in the main body of the text. A useful
reference is {\BDFL}.

\subsec{Construction of ${G_2}$}

\indent ${G_2}$ can be obtained from ${SO}({7})$ as the subgroup leaving
a real spinor, the {\bf 8}, invariant. We will use this fact to give an
explicit construction of ${G_2}$ that also manifests the maximal
${SU}({3})$ subgroup.

\indent We begin with a Majorana representation for the Dirac matrices
of ${SO}({7})$; one explicit choice of real matrices is

$$
{\Gamma^1} =
{\epsilon}{\otimes}{\epsilon}{\otimes}{\epsilon},\,\,{\Gamma^2} =
{1}{\otimes}{\sigma_1}{\otimes}{\epsilon},\,\,{\Gamma^3} = {1}
{\otimes}{\sigma_3}{\otimes}{\epsilon},
$$

$$
{\Gamma^4} =
{\sigma_1}{\otimes}{\epsilon}{\otimes}{1},\,\,{\Gamma^5} =
{\sigma_3}{\otimes}{\epsilon}{\otimes}{1},\,\,{\Gamma^6} =
{\epsilon}{\otimes}{1}{\otimes}{\sigma_1},\,\,{\Gamma^7} =
{\epsilon}{\otimes}{1}{\otimes}{\sigma_3}.\eqno(47)
$$

\noindent Here ${\sigma_i}$ are the usual Pauli matrices and ${\epsilon}
= {i}{\sigma_2}$. It is most convenient to work with a complex basis
${\gamma^a}, {\gamma^{\bar a}}$ for six of the seven matrices, \eg\
$$
{\gamma^1} = {{{i}{\Gamma^1} - {\Gamma^4}}\over{2}}\ ,\ {\gamma^{\bar 1}} =
{{{i}{\Gamma^1} + {\Gamma^4}}\over{2}}\ , {\rm etc.}
$$
and to rename ${\bar\gamma}
= {\Gamma^7}$.

In such a basis, the Dirac algebra $\{\Gamma^I,\Gamma^J\}=-2\delta^{IJ}$
becomes
\eqn\dalg{
\{\gamma^a, \gamma^{\bar b}\} = {\delta^{{a}{\bar b}}}\ ,\ \{{\bar\gamma},
{\gamma^a}\} =\{ {\bar\gamma}, {\gamma^{\bar a}}\} = {0}\ ,\ {\bar\gamma}^2 =
{-1}\ .}
Note also that ${\gamma^{a*}} = {-\gamma^{\bar a}},
{\gamma^{a\dagger}} = {\gamma^{\bar a}}$. The ${\gamma^{a}}$'s and
${\gamma^{\bar a}}$'s behave like fermion creation and
annihilation operators, and we can always find a spinor
${\zeta}$ satisfying

\eqn\fortyeight{
{\gamma^a}{\zeta} = {0}, }

\noindent corresponding to the Fock vacuum. This also implies
\eqn\fortynine{
i{\bar\gamma}{\zeta} = {\zeta}\ ; }
$i{\bar \gamma}$ is analogous to $(-1)^F$.

$G_2$ is the little group leaving a real spinor $\eta$ invariant.
This subgroup is most easily investigated by rotating $\eta$ into the form
\eqn\fifty{
{\eta} = {\zeta} + {\zeta^*} + {\zeta^c} + {\zeta^{c*}}, }
where
\eqn\fiftyone{
{\zeta^c} = {\gamma^{\bar 1}}{\gamma^{\bar 2}}{\gamma^{\bar 3}}{\zeta}.
}
Note that

\eqn\fiftytwo{
{\gamma^{\bar a}}{\zeta^c} = {0}. }

\noindent The generators of ${SO}({7})$ are the 21 real combinations of

\eqn\fiftythree{
{\gamma^{ab}}\ ,\ {\gamma^{{\bar a}{\bar b}}}\ ,\ {\bar\gamma}{\gamma^a}\
,\
{\bar\gamma}{\gamma^{\bar a}}\ ,\quad {\rm and}\quad
{\gamma^{{a}{\bar b}}}. }

\noindent The six generators ${\gamma^{{a}{\bar b}}}, {a}{\not=}{b}$,
automatically annihilate ${\eta}$. One can also easily show that
the two generators

\eqn\fiftyfour{
{\gamma^{{1}{\bar 1}}} - {\gamma^{{2}{\bar 2}}}, {\gamma^{{2}{\bar 2}}}
- {\gamma^{{3}{\bar 3}}} }

\noindent annihilate ${\eta}$, as do the six generators

\eqn\exgen{
{\bar\gamma}{\gamma^a} + \frac{i}{2} {\epsilon^{abc}}\gamma^{
{\bar b}{\bar c}}\ ,\
{\bar\gamma} {\gamma^{\bar a}} {+} \frac{i}{2}
{\epsilon^{abc}}\gamma^{bc}\ .
}

\noindent The fourteen real combinations of these fourteen matrices
generate ${G_2}$.

The ${SU}({3})$ subgroup is easily exhibited. First choose a basis for
the 7 dimensional spinor subspace orthogonal to ${\eta}$:

\eqn\sbasis{
{\gamma^a}{\eta}\ ,\ {\gamma^{\bar a}}{\eta}\ ,\ {\bar\gamma}{\eta}. }

\noindent These give a basis for the {\bf 7} of ${G_2}$. Working in this
basis, the real generators ${\gamma^{{a}{\bar b}}} + {\gamma^
{{\bar a}{b}}}$, ${i} ({\gamma^{{a}{\bar b}}} - {\gamma^{{\bar a}{b}}})$,
(with $a\neq b$)
and ${i} ({\gamma^{{a}{\bar a}}} - {\gamma^{{b}{\bar b}}})$ correspond to
matrices of the form

$$
{}\pmatrix{\lambda&0&0\cr
0&-\lambda^T&0\cr
0&0&0\cr}
$$

\noindent where ${\lambda}$ are ${3}{\times}{3}$ generators of
${SU}({3})$. Thus these generators give the ${SU}({3})$ subgroup, and
the decomposition ${\bf 7} = {\bf 3} + {\bf \bar 3} + {\bf 1}$ is
manifest. One can also easily work out the explicit matrix
representations of the remaining 6 generators \exgen\ of ${G_2}$.

\subsec{Invariants of ${G_2}$}

\indent As shown in table I, ${G_2}$ has a fully antisymmetric invariant
${f^{\alpha\beta\gamma}}$. this can easily be obtained from the spinor
${\eta}$:

\eqn\fiftyfive{
{f^{\alpha\beta\gamma}} = {\eta^T}{\Gamma^{\alpha\beta\gamma}}{\eta}, }

\noindent with normalization ${\eta}^T\eta=1$. Using the preceding
construction of ${\eta}$, the ${f^{\alpha\beta\gamma}}$'s can be explicitly
computed.

\indent One cannot form lower ${G_2}$ invariants, such as
${\eta^T}{\Gamma^\alpha}{\eta}$ or ${\eta^T}{\Gamma^{\alpha\beta}}{\eta}$,
because
of the antisymmetry of ${\Gamma^\alpha}, {\Gamma^{\alpha\beta}}$.
The invariants ${\eta^T}{\Gamma^{\alpha\beta\gamma\delta}}{\eta}$,
$\cdots$,
${\eta^T}{\Gamma^{{\alpha_1}{\cdots}{\alpha_7}}}{\eta}$ are
trivially duals of
the lower invariants.

\indent At first sight it would appear that there are many other
invariants that can be constructed from the primitives
${\delta^{\alpha\beta}}$, ${f^{\alpha\beta\gamma}}$, and
${\epsilon^{{\alpha_1}
{\cdots}{\alpha_7}}}$ by contracting products. However, the primitives
satisfy a number of relations that restrict the number of possible
invariants. A useful starting point is the Fierz identity

\eqn\Fierz{
{\eta^T}{\gamma^{\alpha\beta\gamma}}{\eta}\,{\eta^T}{\gamma^
{\gamma\delta\epsilon}}
{\eta} = \frac{1}{8} {\eta^T}
{\gamma^{\alpha\beta\gamma}} {\gamma^{\gamma\delta
\epsilon}} {\eta} + \frac{1}{48}
{\eta^T}{\gamma^{\phi \eta \kappa}} {\eta}\,{\eta^T}
\gamma^{\alpha\beta\gamma}{\gamma^{\phi
\eta \kappa}}{\gamma^{\gamma\delta\epsilon}} {\eta}. }

\noindent From this one can show that the totally antisymmetrized
product of two ${f}$'s is equivalent to the dual of ${f}$,

\eqn\fiftysix{
{f^{[\alpha\beta\gamma}}{f^{\gamma\delta\epsilon]}} = {\tilde f}^{\alpha
\beta\delta\epsilon}, }

\noindent where on the left we antisymmetrize on
${\alpha\beta\delta\epsilon}$ and on the right ${\tilde f}$ is given by

\eqn\fiftyseven{
{\tilde f}^{\alpha\beta\gamma\delta} = \frac{1}{3!} {\epsilon^{\alpha
\beta\gamma\delta\epsilon\phi\eta}} {f^{\epsilon\phi\eta }}. }

\noindent Likewise, one can prove the identity \refs{\BDFL}

\eqn\fiftyeight{
{f^{\alpha\beta\gamma}}{f^{\alpha\delta\epsilon}} +
{f^{\alpha\delta\gamma}} {f^{\alpha\beta\epsilon}} =
{2\delta^{\beta\delta}}{\delta^{\gamma\epsilon}} {-}
{\delta^{\gamma\delta}}{\delta^{\beta\epsilon}} {-}
{\delta^{\beta\gamma}}{\delta^{\delta\epsilon}} }

\noindent which shows that the remaining components of the product of
two ${f}$'s are not independent. The triple identity
\eqn\fiftynine{
\eqalign{
{f^{\alpha\beta\gamma}}{f^{\gamma\delta\epsilon}}{f^{\epsilon\phi\eta}}=&
{\delta^{\delta\alpha}}{f^{\beta\phi\eta}} +
{\delta^{\phi\alpha}}{f^{\beta\eta\delta}} + \delta^{\alpha\eta}
f^{\beta\delta\phi}
+{\delta^{\delta\eta}}{f^{\alpha\beta\phi}}\cr {-}
&{\delta^{\delta\beta}}{f^{\alpha\phi\eta}} {-}
{\delta^{\beta\phi}}{f^{\alpha\eta\delta}}
{-} {\delta^{\eta\beta}}{f^{\alpha\delta\phi}} {-}
{\delta^{\delta\phi}}{f^{\alpha\beta\eta}}\cr }}
also follows, and shows that any higher product is not
independent.

\indent Finally, an important relation comes from
the observation that the fully antisymmetrized product
${f^{[\alpha\beta\gamma}{\tilde f}^{\delta\epsilon\phi\eta]}}$ must be
proportional to ${\epsilon^{\alpha\beta\gamma\delta\epsilon\phi\eta}}$.
Contracting both with ${\epsilon^{\alpha\beta\gamma\delta\epsilon\phi
\eta}}$ and using
$$
f^{\alpha\beta\gamma} f^{\alpha\beta\gamma} = 7\cdot 6
$$
(from \fiftyeight) shows that
\eqn\epsident{
{\epsilon^{\alpha\beta\gamma\delta\epsilon\phi\eta}} = {5}\,\,
{f^{[\alpha\beta\gamma}}{\tilde f}^{\delta\epsilon\phi\eta]}\ . }
As a result, the invariants ${\delta^{\alpha\beta}},
{f^{\alpha\beta\gamma}}$ and ${\tilde f}^{\alpha\beta\gamma\delta}$ can
be taken to be the only independent invariants. Other invariant tensors
can always be reduced to products of these.

\indent The identity \epsident\ can be used to prove identities relating
mesons and baryons.
Substituting \epsident\ into the relation\foot{In subsequent formulas
the height of indices is only significant for its convenience.}
\eqn\sixty{
{3!}{\delta}^{{\alpha^{\prime}}{\beta^{\prime}}{\gamma^{\prime}}
{\delta^{\prime}}}_{\alpha\beta\gamma\delta } = {\epsilon}^
{{\alpha^{\prime}}{\beta^{\prime}}{\gamma^{\prime}}{\delta^{\prime}}{\mu}
{\nu}{\lambda}} {\epsilon_{\alpha\beta\gamma\delta\mu\nu\lambda}} }

\noindent leads to the identity

\eqn\Newident{\eqalign{0=&
{7}{\delta}^{{\alpha^{\prime}}{\beta^{\prime}}{\gamma^{\prime}}
{\delta^{\prime}}}_{\alpha\beta\gamma\delta } + {3}{\tilde f}_
{\alpha\beta\gamma\delta} {\tilde f}^{{\alpha^{\prime}}{\beta^{\prime}}
{\gamma^{\prime}}{\delta^{\prime}}} + {32} {\delta}^{[{\alpha^{\prime}}}
_{[\alpha} {f}^{{\beta^{\prime}}{\gamma^{\prime}}
{\delta^{\prime}}]}{f}_{\beta
\gamma\delta]}\cr
&{-} {16} {\delta}^{[{\alpha^{\prime}}}
_{[\alpha} {\tilde f}^{{\beta^{\prime}}{\gamma^{\prime}}{\delta^{\prime}}]
{\epsilon}}{\tilde f}_{\beta\gamma\delta]\epsilon} {-} {72} {\delta}
^{[{\alpha}}_{[{\alpha}}{\delta}^{\beta^{\prime}}_\beta
{f}^{{\gamma^{\prime}}{\delta^{\prime}}]},_{\gamma\delta]} \cr}}
where
$$
f_{\alpha\beta,\gamma\delta}= f_{\alpha\beta\epsilon}
f_{\gamma\delta\epsilon}\ .
$$
This can be simplified into a form more useful for
proving proving the constraints in Section 3. Using two relations
derived from the triple identity (by contracting it with an
${f_{\mu\nu\lambda}}$ on one and two indices)
$$
{\tilde f}^{{\beta^{\prime}}{\gamma^{\prime}}{\delta^{\prime}}{\epsilon}}
{\tilde f}_{\beta\gamma\delta\epsilon} = {3} ({\delta}^{[{\beta^{\prime}}}
_{[\beta} {f}^{{\gamma^{\prime}}{\delta^{\prime}}]},_{{\gamma}{\delta}]}
{-} {f}^{[{\beta^{\prime}}{\gamma^{\prime}}}_{[\beta}{f}_{\gamma\delta]}
^{{\delta^{\prime}}]})
$$
\noindent and
$${f}^{{\gamma^{\prime}}{\delta^{\prime}}},_{\gamma\delta}
={\tilde f}^{{\gamma^{\prime}}{\delta^{\prime}}}_{\gamma\delta} +
{\delta}^{{\gamma^{\prime}}{\delta^{\prime}}}_{\gamma\delta},$$
we have
\eqn\fident{\eqalign{0=&
{\tilde f}_{\alpha\beta\gamma\delta}{\tilde
f}^{{\alpha^{\prime}}{\beta^{\prime}}{\gamma^{\prime}}{\delta^{\prime}}}
+ \frac{32}{3} {\delta}^{[{\alpha^{\prime}}}_{[{\alpha}}{f}
^{{\beta^{\prime}}{\gamma^{\prime}}{\delta^{\prime}}]}
{f}_{\beta\gamma\delta]} + {16}
{\delta}^{[{\alpha^{\prime}}}_{[{\alpha}} {f_\beta}^{{\beta^{\prime}}
{\gamma^{\prime}}} {f}_{\gamma\delta]}^{{\delta^{\prime}}]}\cr
&{-} {40}{\delta}^{[{\alpha'}}_{[{\alpha}}{\delta}^{\beta^{\prime}}_\beta
{\tilde f}^{{\gamma^{\prime}}{\delta^{\prime}}]}_{\gamma\delta]} {-}
{\delta}^{{\alpha^{\prime}}{\beta^{\prime}}
{\gamma^{\prime}}{\delta^{\prime}}}
_{\alpha\beta\gamma\delta}} }
This identity implies the constraints
\thirtyfour, \fortytwo\ relating mesons and baryons.

\listrefs

\end